\documentclass[journal=jacsat,manuscript=article]{achemso}

\usepackage[version=3]{mhchem} 
\usepackage{subcaption}
\usepackage{pdfpages}
\usepackage{algpseudocode} 
\usepackage{siunitx}
\usepackage[section]{placeins}
\usepackage{multirow}
\usepackage{txfonts}
\usepackage{amsmath}
\usepackage{hhline}
\usepackage{caption}
\usepackage{algorithm2e}
\usepackage{soul}

\newcommand{\affa}{Sorbonne Université, Laboratoire de Chimie Théorique(UMR-7616-CNRS), 4 place Jussieu-75005 Paris, France}

\newcommand{\affb}{Sorbonne Université, CNRS, Université Paris Cité, Laboratoire Jacques Louis Lions (LJLL), 4 place Jussieu-75005 Paris, France}

\newcommand{\affc}{TotalEnergies, Tour Coupole La D\'{e}fense, 2 Pl. Jean Millier, 92078 Paris, France}

\newcommand{\affe}{Institut Universitaire de France, Paris, France}

\author{Mohammad Haidar}
\affiliation{\affa}
\alsoaffiliation{\affb}
\alsoaffiliation{\affc}

\author{Marko J. Rančić}
\affiliation{\affc}

\author{Yvon Maday}
\affiliation{\affb}
\alsoaffiliation{\affe}

\author{Jean-Philip Piquemal}
\affiliation{\affa}

\email{Mohammadhaidar2016@outlook.com,jean-philip.piquemal@sorbonne-universite.fr}

\title[An \textsf{achemso} demo]
  {Extension of the Trotterized Unitary Coupled Cluster to Triple Excitations}

\abbreviations{IR,NMR,UV}

\begin{document}


\begin{abstract}
The Trotterized Unitary Coupled Cluster Single and Double (UCCSD) ansatz has recently attracted interest due to its use in Variation Quantum Eigensolver (VQE) molecular simulations on quantum computers. However, when the size of molecules increases, UCCSD becomes less interesting 
as it cannot achieve sufficient accuracy. Including higher-order excitations is therefore mandatory to recover the UCC's missing correlation effects. In this Letter, we extend the Trotterized UCC approach via the addition of (true) Triple T excitations introducing UCCSDT. We also include both spin and orbital symmetries. Indeed, in practice, these later help to reduce unnecessarily  circuit excitations and thus accelerate the optimization process enabling to tackle larger molecules.  Our initial numerical tests (12-14 qubits) show that UCCSDT improves the overall  accuracy by at least two-orders of magnitudes with respect to standard UCCSD. Overall, the UCCSDT ansatz is shown to reach chemical accuracy and to be competitive with the CCSD(T) gold-standard classical method of quantum chemistry.
\end{abstract}

\section{Introduction}
The use of quantum computers is a promising strategy to overcome challenges occuring in the area of quantum chemistry.\cite{aspuru2005simulated,reiher2017elucidating,bauer2020quantum,mcardle2020quantum,bassman2021simulating}.  Such hardware potentially facilitate the encoding of the full configuration interaction (FCI) wavefunction of the many-electron molecular system thanks to entangled quantum bits or qubits.  In classical computers, the FCI includes a number of determinants which scales exponentially with the number of electrons denoted $n_e$, as roughly $O(n_o^{n_e})$, where $n_o$ is the number of spin-orbitals, which makes the manipulation and storage of the wavefunction inefficient\cite{gan2006lowest,lehtola2017cluster,vogiatzis2017pushing}. However, using a quantum computer, we can instead store the (FCI) wavefunction by using only $n_o$ orbitals which corresponds to $n_o$ qubits\cite{aspuru2005simulated}. This potential quantum advantage has recently excited both hardware and software communities generating rapid progresses in the field. Several quantum algorithms have been developed, and among them, the Variational Quantum Eigensolver (VQE)\cite{peruzzo2014variational,mcclean2016theory,cerezo2021variational,li2022toward} appears well suited for its practical implementation on present Noisy Intermediate Scaled Quantum (NISQ) devices
\cite{o2016scalable,bharti2022noisy}. VQE simulations have been performed numerically for molecules using various virtual noiseless simulators\cite{kuhn2019accuracy,lolur2021benchmarking,yeter2021benchmarking,cao2021towards} and  have been tested experimentally on actual NISQ devices\cite{peruzzo2014variational,google2020hartree,kandala2017hardware,hempel2018quantum,PRXQuantum.3.040318}. An important 
 component in VQE is the parametrized ansatz, which represents the trial wavefunction and  is implemented as a quantum circuit, composed of unitary quantum gates, that  measures the energy expectation value of the trial wavefunction and then its parameters are updated in a classical optimization loop. 
The Unitary Coupled-Cluster (UCC) ansatz was used when VQE was initially proposed, and has granted lots of attention since then. Indeed, UCC includes several attractive features: (i) it is chemistry-inspired since it is a unitary version of the classical coupled-cluster method, which is
among the most accurate classical quantum chemistry methods
for many-body simulation (see exhaustive recent reviews in \cite{crawford2007introduction,bartlett2007coupled}); (ii) A variant of the UCC theory including single (S) and double (D) excitations called UCCSD  is well suited for a direct use on the current NISQ circuits (see comprehensive reviews of UCCSD in \cite{romero2018strategies,anand2022quantum}). Moreover, the Suzuki-Trotter approximation\cite{hatano2005finding} allows separating out the ansatz into a product form  of individual exponentials. This version of the ansatz is then denoted Trotterized UCC. Since it is unitary, it can 
 be easily transformed into well-defined unitary gates and therefore its implementation is possible on NISQ devices.  

The Trotterized  UCCSD ansatz is well documented  \cite{romero2018strategies,anand2022quantum,sokolov2020quantum,xia2020qubit,grimsley2019trotterized} and has been successfully tested for small molecules\cite{yung2014transistor,shen2017quantum,o2022purification}. However, UCCSD exhibits a limited accuracy in strongly correlated systems similarly as the classical CCSD ansatz\cite{xia2020qubit,mizukami2020orbital}. Moreover, when the size of the molecules increases, many unnecessary excitations can appear in UCCSD which do not improve the chemical accuracy while adding extra-noise due to the extra circuit-depth.\cite{Benzenewassil}
This reveals the need for more accurate and compact representations of the wavefunction. In that context, various alternative versions of the UCC ansatz, also truncated at the double excitations level have been proposed by the community and  
 have brought interesting upgrades\cite{grimsley2019trotterized,lee2018generalized,kohn2022capabilities,grimsley2019adaptive,fedorov2022unitary}. A possible direction also is to use the contracted quantum eigensolvers (CQE) methods which provides an ansatz that can generalize CCSD towards FCI\cite{smart2022resolving,smart2021quantum}.
However, a possible direction towards improving accuracy is to extend the Trotterized UCCSD  through adding three-body clusters that represent  the (true) Triple T excitations. For example, the gold-standard CCSD(T) method\cite{raghavachari2013historical} of (classical) quantum chemistry involves perturbative triple excitations that play a key role in the inclusion of correlation effects \cite{ramabhadran2013extrapolation}.\\
The objective of this Letter is to introduce the
Trotterized UCCSDT approach and to analyze the behaviour of triple excitations on a set of molecules compared to the initial UCCSD. Of course, an expected downside of the UCCSDT approach is that it involves a large number of parameters to optimize and therefore leads to generate circuits that are excessively long and  therefore  not suited for their practical use on current NISQ devices. A possible way to cope with this potential  inefficiency  of UCCSDT (i.e allowing to reduce the required number of parameters) is to use the power of the total spin and point group symmetries. Indeed, taking into account these symmetries enable to use spin
factorization and orbital symmetry techniques\cite{crawford2007introduction} that have been shown to enable simplifications in the classical coupled cluster
equations through eliminating unnecessarily single and double excitations.
Very recently, the point group symmetry constraint has been implemented with the UCCSD ansatz\cite{anand2022quantum,cao2021towards}. The authors demonstrated a significant reduction of the number of parameters without affecting the accuracy.\\
This article is organized as follows:
the next section is devoted to a description of the theory of Unitary Coupled Cluster method for electronic structure calculations. It includes the detailed formalism of triple excitations. It is given in its simplified form after application of both spin and orbital symmetries. The subsequent section presents the computational materials used to describe the UCCSDT-VQE approach and the last section illustrates the application of UCCSDT-VQE by showing numerical results through testing several molecules. Comparisons were made with UCCSD as well as the classical methods such as CCSD, CCSD(T), CCSDT-full and  FCI.

\section{Theory}
In this work we have based ourselves on the Born-Oppenheimer clamped-nuclei Hamiltonian formalism:
\begin{equation}
\label{Hamiltonian}
    \hat{H} = \sum_{p,q} h_{pq} \hat{\textrm{a}}_p^{\dagger} \hat{\textrm{a}}_q + \frac{1}{2} \sum_{p,q,r,s} h_{pqrs} \hat{\textrm{a}}_p^{\dagger} \hat{\textrm{a}}_q^{\dagger} \hat{\textrm{a}}_r \hat{\textrm{a}}_s
\end{equation}
where $a_p^{\dagger}$ ($a_q$) are anti-commuting operators that create (annihilate) electrons in molecular spin-orbital $p(q)$, respectively.
The symbols $h_{pq}$ and $h_{pqrs}$ denote the one- and two-body integrals of the corresponding operators and spin-orbitals in Dirac notation, respectively. These integrals can be easily computed on a classical computer.\\
The main goal is to find the ground state energy $E_0$ of  Hamiltonian $\hat{H}$ given in eq. \eqref{Hamiltonian}. The Variational Quantum Eigensolver (VQE),  a hybrid quantum classical
algorithm was designed to solve this problem (see Figure 1 in \cite{peruzzo2014variational}). The principal of VQE is that the quantum computer prepares and measures, for any given parameter $\vec{\theta}$, the parametrized quantum state  $|\Psi(\vec{\theta})\rangle$ while the classical computer is used for optimization to propose and update $\vec{\theta}$, in order to minimize the variational energy
 $\langle \Psi(\vec{\theta})| H |\Psi(\vec{\theta})\rangle \ge E_0$.\\
 We focus on the Unitary Coupled Cluster (UCC) wavefunction, which, as mentioned in the introduction, is chemically inspired from the classical Coupled Cluster (CC) method, that has yielded high-accuracy results in classical
quantum chemistry. The UCC theory is important in our study, since its wavefunction is unitary and therefore can be efficiently implemented on the quantum computer unlike the regular CC method. The UCC wavefunction can be expressed as follows:
\begin{equation}
\label{exponential}
|\Psi(\vec{\theta})\rangle = e^{\hat{T}(\vec{\theta}) - \hat{T}^{\dagger}(\vec{\theta})}| \psi_\mathrm{HF}\rangle
\end{equation}
where $\hat{T}$ is the so-called cluster excitation operator and $\hat{T}^{\dagger}$ is its Hermitian conjugate. $|\psi_\mathrm{HF}\rangle$ is the reference Hartree-Fock state.  Since we are interested in investigating the effect of Triple excitations on UCC's accuracy, the coupled-cluster
excitations $\displaystyle \hat{T}(\vec{\theta})$ are truncated to single, double and triple excitations to introduce UCCSDT,
\begin{equation}
\label{nexcitations}
 \hat{T}(\vec{\theta}) = \hat{T_1}(\vec{\theta}) + \hat{T_2}(\vec{\theta})+ \hat{T_3}(\vec{\theta}) = \sum_{a,i} \theta_{i}^{a} \,\hat{\textrm{a}}_a^\dagger \hat{\textrm{a}}_i  +  \sum_{a,b,i,j}\theta_{i,j}^{a,b}\,\hat{\textrm{a}}_a^\dagger \hat{\textrm{a}}_b ^\dagger \hat{\textrm{a}}_i \hat{\textrm{a}}_j  + 
 \sum_{a,b,c,i,j,k}\theta_{i,j,k}^{a,b,c}\, \hat{\textrm{a}}_a^\dagger \hat{\textrm{a}}_b ^\dagger \hat{\textrm{a}}_c ^\dagger \hat{\textrm{a}}_i \hat{\textrm{a}}_j \hat{\textrm{a}}_k
\end{equation}
 where $a,b,c \in virt, i,j,k \in occ$, $\theta_{i}^{a}$, $\theta_{i,j}^{a,b}$ and $\theta_{i,j,k}^{a,b,c}$ are 
 the variational parameters. The triple excitation term  $\hat{T_3}(\vec{\theta})$ (given in equation \eqref{nexcitations}) increases 
 drastically the number of parameters as $n_o$ increases. This makes the optimization process very difficult and slow to reach convergence. However, as discussed in the introduction, the many excitations appearing in the UCCSDT wavefunction (single, double and triple) are not all important and will not affect all the correlation energy.  Therefore, determining, {\sl a priori}, those unimportant amplitudes, and thus not including them in equation \eqref{nexcitations} above
before compilation and execution of a variational optimization of the parameters, is highly significant  to save resources. In this work we focus only on neutral closed-shell molecular species allow to make the electronic calculations on such molecules very efficient by exploiting simultaneously both spin and orbital
symmetries in the UCCSDT ansatz. Indeed, since in the neutral closed-shell molecules exhibit an even number of electrons, one can maintain only excitations that can satisfy spin symmetry through keeping the balance between spin-up($\uparrow$)  and spin-down($\downarrow$)
electrons. Therefore it is straightforward to include the spin symmetry  constraint in equation \eqref{nexcitations}. The inclusion of orbital symmetry constraints requires more precise explanation for its implementation for each molecule subject to a non-trivial point group symmetry \cite{dresselhaus2007group,tinkham1964group}. For example the H$_2$ and N$_2$ molecules belongs to 
 the $D_{\infty h}$ point group whereas LiH and H$_2$O  belong to $C_{\infty v}$. Each such group is characterized by its irreducible
representations (irreps) which are listed in \cite{Tablesgroup}.
Let us now describe how to impose the orbital symmetry constraint from each irreps group into equation \eqref{nexcitations} following  the rules  that are given in chapter 
 2 of reference \cite{crawford2007introduction}, and that were recently reviewed in \cite{anand2022quantum}. First, we use the orbital symmetry operator  $\hat{s}_e$, noted in \cite{anand2022quantum}, which maps 
\begin{itemize}
\item each orbital to the irreps of that orbital
\item each pair of orbitals to the tensor product of the two irreps of the two orbitals
\item each triple of orbitals to the tensor product of the three irreps of the three orbitals 
\end{itemize}
Assuming e.g. that the $D_{2 h}$ group is used, and taking for example the triple excitation case, if the irreps of  $i,j,k$  are  $i = A_u, j=B_{1g},k= B_{2u}$ ,  then we have : $\hat{s}_e(i) =A_u $, $\hat{s}_e(j) =B_{1g} $, $\hat{s}_e(k) = B_{2u}$  and  $\hat{s}_e(i,j,k) = B_{3g}$ according to the multiplication table for $D_{2 h}$. Then, the complexity reduction comes from the rule that single excitations must only occur between  orbitals that belong to the same irreps. Double and Triple excitations must only occur between occupied and virtual orbitals whose tensor product irreps are similar. Furthermore, the products of irreps that are partaken in each excitation must contain the fully symmetric irreps.

 Merging these constraints with the spin symmetry, we obtain the operator for Singles (S)
\begin{equation}
\label{single}
\hat{T}_1(\vec{\theta}) - \hat{T}^{\dagger}_1(\vec{\theta})= \sum_{\substack{i,a \in \{\uparrow\} \\ \hat{s}_e(i)=\hat{s}_e(a)}} \theta_i^a(\hat{\textrm{a}}_a^{\dagger} \hat{\textrm{a}}_i - \hat{\textrm{a}}_i^{\dagger} \hat{\textrm{a}}_a)
+ \sum_{\substack{\bar{i},\bar{a} \in \{\downarrow\} \\ \hat{s}_e(\bar{i})=\hat{s}_e(\bar{a})}} \theta_{\bar{i}}^{\bar{a}}(\hat{\textrm{a}}_{\bar{a}}^{\dagger} \hat{\textrm{a}}_{\bar{i}} - \hat{\textrm{a}}_{\bar{i}}^{\dagger} \hat{\textrm{a}}_{\bar{a}})
\end{equation}

For Doubles (D),
\begin{equation}
\label{double}
\begin{split}
\hat{T}_2(\vec{\theta}) - \hat{T}^{\dagger}_2(\vec{\theta}) &= \sum_{\substack{i<j,a<b \in \{\uparrow\} \\ \hat{s}_e(i,j)=\hat{s}_e(a,b)}} \theta_{i,j}^{a,b}(\hat{\textrm{a}}_a^{\dagger} \hat{\textrm{a}}_b^{\dagger} \hat{\textrm{a}}_j \hat{\textrm{a}}_i - \hat{\textrm{a}}_i^{\dagger} \hat{\textrm{a}}_j^{\dagger} \hat{\textrm{a}}_b \hat{\textrm{a}}_a) + \sum_{\substack{\bar{i}<\bar{j},\bar{a}<\bar{b} \in \{\downarrow\} \\ \hat{s}_e(\bar{i},\bar{j})=\hat{s}_e(\bar{a},\bar{b})}} \theta_{\bar{i},\bar{j}}^{\bar{a},\bar{b}}(\hat{\textrm{a}}_{\bar{a}}^{\dagger} \hat{\textrm{a}}_{\bar{b}}^{\dagger} \hat{\textrm{a}}_{\bar{j}} \hat{\textrm{a}}_{\bar{i}} - \hat{\textrm{a}}_{\bar{i}}^{\dagger} \hat{\textrm{a}}_{\bar{j}}^{\dagger} \hat{\textrm{a}}_{\bar{b}} \hat{\textrm{a}}_{\bar{a}})\\
&+\sum_{\substack{i,\bar{j},a,\bar{b} \\ i,a \in \{\uparrow\}; \bar{j},\bar{b} \in \{\downarrow\} \\ \hat{s}_e(i,\bar{j})=\hat{s}_e(a,\bar{b})}} \theta_{i,\bar{j}}^{a,\bar{b}}(\hat{\textrm{a}}_{a}^{\dagger} \hat{\textrm{a}}_{\bar{b}}^{\dagger} \hat{\textrm{a}}_{\bar{j}} \hat{\textrm{a}}_{i} - \hat{\textrm{a}}_{i}^{\dagger} \hat{\textrm{a}}_{\bar{j}}^{\dagger} \hat{\textrm{a}}_{\bar{b}} \hat{\textrm{a}}_{a})
\end{split}
\end{equation}

For Triples (T),
\begin{equation}
\label{Triple}
\begin{split}
\hat{T}_3(\vec{\theta}) - \hat{T}^{\dagger}_3(\vec{\theta}) &= \hspace{-0.4cm} \sum_{\substack{i<j<k \in \{\uparrow\} \\ a<b<c \in \{\uparrow\} \\ \hat{s}_e(i,j,k)=\hat{s}_e(a,b,c)}} \hspace{-0.3cm} \theta_{i,j,k}^{a,b,c}(\hat{\textrm{a}}_a^{\dagger} \hat{\textrm{a}}_b^{\dagger} \hat{\textrm{a}}_c^{\dagger} \hat{\textrm{a}}_k \hat{\textrm{a}}_j \hat{\textrm{a}}_i - \hat{\textrm{a}}_i^{\dagger} \hat{\textrm{a}}_j^{\dagger} \hat{\textrm{a}}_k^{\dagger}  \hat{\textrm{a}}_c \hat{\textrm{a}}_b \hat{\textrm{a}}_a) + \hspace{-0.4cm} \sum_{\substack{\bar{i}<\bar{j}<\bar{k} \in \{\downarrow\} \\ \bar{a}<\bar{b}<\bar{c} \in \{\downarrow\} \\ \hat{s}_e(\bar{i},\bar{j},\bar{k})=\hat{s}_e(\bar{a},\bar{b},\bar{c})}} \hspace{-0.3cm} \theta_{\bar{i},\bar{j},\bar{k}}^{\bar{a},\bar{b},\bar{c}}(\hat{\textrm{a}}_{\bar{a}}^{\dagger} \hat{\textrm{a}}_{\bar{b}}^{\dagger} \hat{\textrm{a}}_{\bar{c}}^{\dagger} \hat{\textrm{a}}_{\bar{k}} \hat{\textrm{a}}_{\bar{j}} \hat{\textrm{a}}_{\bar{i}} - \hat{\textrm{a}}_{\bar{i}}^{\dagger} \hat{\textrm{a}}_{\bar{j}}^{\dagger} \hat{\textrm{a}}_{\bar{k}}^{\dagger} \hat{\textrm{a}}_{\bar{c}} \hat{\textrm{a}}_{\bar{b}} \hat{\textrm{a}}_{\bar{a}})\\
&+ \hspace{-0.4cm} \sum_{\substack{i<j,a<b \in \{\uparrow\} \\ \bar{k},\bar{c} \in \{\downarrow\} \\ \hat{s}_e(i,j,\bar{k})=\hat{s}_e(a,b,\bar{c})}} \hspace{-0.3cm} \theta_{i,j,\bar{k}}^{a,b,\bar{c}}(\hat{\textrm{a}}_a^{\dagger} \hat{\textrm{a}}_b^{\dagger} \hat{\textrm{a}}_{\bar{c}}^{\dagger} \hat{\textrm{a}}_{\bar{k}} \hat{\textrm{a}}_j \hat{\textrm{a}}_i - \hat{\textrm{a}}_i^{\dagger} \hat{\textrm{a}}_j^{\dagger} \hat{\textrm{a}}_{\bar{k}}^{\dagger}  \hat{\textrm{a}}_{\bar{c}} \hat{\textrm{a}}_b \hat{\textrm{a}}_a) + \hspace{-0.4cm} \sum_{\substack{\bar{i}<\bar{j},\bar{a}<\bar{b} \in \{\downarrow\} \\ k,c \in \{\uparrow\} \\ \hat{s}_e(i,j,\bar{k})=\hat{s}_e(a,b,\bar{c})}} \hspace{-0.3cm} \theta_{\bar{i},\bar{j},k}^{\bar{a},\bar{b},c}(\hat{\textrm{a}}_{\bar{a}}^{\dagger} \hat{\textrm{a}}_{\bar{b}}^{\dagger} \hat{\textrm{a}}_c^{\dagger} \hat{\textrm{a}}_k \hat{\textrm{a}}_{\bar{j}} \hat{\textrm{a}}_{\bar{i}} - \hat{\textrm{a}}_{\bar{i}}^{\dagger} \hat{\textrm{a}}_{\bar{j}}^{\dagger} \hat{\textrm{a}}_k^{\dagger}  \hat{\textrm{a}}_c \hat{\textrm{a}}_{\bar{b}} \hat{\textrm{a}}_{\bar{a}})
\end{split}
\end{equation}
With these  
excitations 
 we construct the symmetric UCCSDT, which is named throughout the text as sym-UCCSDT.
\section{Computational Procedure}


%
%
%

The implementation of an exponential operator $e^{\hat{T}(\vec{\theta}) - \hat{T}^{\dagger}(\vec{\theta})}$ given in eq. \eqref{exponential} cannot be  done directly on quantum hardware.
Therefore applying Trotterization using the Suzuki–Trotter
expansion\cite{hatano2005finding} is  mandatory to break up the exponential of a sum as a product of individual exponentials. Thus we used this approach to write the Trotterized   $e^{\hat{T}(\vec{\theta}) - \hat{T}^{\dagger}(\vec{\theta})}$ with  $T(\theta)-T(\theta)^\dagger$ (eq. \eqref{nexcitations})  in the form:
\begin{equation} 
\label{tettor}
e^{\hat{T}(\vec{\theta}) - \hat{T}^{\dagger}(\vec{\theta})} =
\bigg(\prod_{\rho} e^{\frac{\theta_{\rho}}{t} (T_\rho-T_\rho^\dagger)}\bigg)^t  + \mathcal{O}\,(\frac{1}{t})
\end{equation} 
where $t$ is the number of  Trotter steps and $\rho$ corresponds to the elements of excitations introduced in equation \eqref{nexcitations}.
The essential tools needed to perform the VQE process (such as fermionic second quantization of $\hat{H}$ equation \eqref{Hamiltonian}, fermion-qubit
transforms  using Jordan-Wigner representation\cite{fradkin1989jordan}, Trotterization, etc) are implemented in the myQLM-fermion package\cite{myqlm-fermion,haidar2022open}. We use the UCC family module from our OpenVQE package\cite{openvqe,haidar2022open} to implement UCCSDT excitations (equations~\eqref{single}~--~\eqref{Triple}). 
We also used PySCF package\cite{sun2018pyscf} 
 to evaluate the molecular orbital integrals $h_{pq}$ and $h_{pqrs}$ which are given in equation \eqref{Hamiltonian}. Using PySCF package, we determine automatically the irreps of each molecular orbital as well as the direct tensor product of the irrep
of the molecular spin-orbitals by using the product
table\cite{Tablesgroup}. Finally, 
we used a gradient-based optimization
method (BFGS) 
 from \textit{scipy.optimize} library\cite{scipy} for  optimizing  the variational parameters  $\theta_{i}^{a}$, $\theta_{i,j}^{a,b}$ and $\theta_{i,j,k}^{a,b,c}$.   The gradient norm
used is equal to 10$^{-10}$. \\
For the calculation of classical chemistry methods CCSD and CCSD(T), we use PYSCF package. However for the full CCSDT, we use CCT3\cite{CCT3} plugin that is provided in Psi4 package\cite{smith2020psi4}.  Given that CC(t;3) method is supposed to yield practically as accurate as full CCSDT for total electronic energies. The energy convergence
threshold used is 10$^{-10}$ Ha.  
\section{Numerical Results and Discussions}
We tested our UCCSDT-VQE (eq. \eqref{nexcitations}) and sym-UCCSDT-VQE (eqs.~\eqref{single}~--~\eqref{Triple}) methods on three sets of molecules: LiH, BeH$_2$ and H$_2$O.  For sym-UCCSD and sym-UCCSDT-VQE calculations, we chose  the highest possible Abelian point groups that belong to each of the systems i.e by refereeing to \cite{DATAbase} we have  $C_{2v}$ point group for LiH and H$_2$O but D$_{2h}$ point group for BeH$_2$. 
The minimal STO-3G basis\cite{hehre1970self} is used in order to minimize the computational costs. Within the STO-3G basis set, LiH  has a Hilbert space spanned by 12 Hartree-Fock (HF) orbitals (4 occupied, 8 virtual). BeH$_2$  has 14 HF orbitals (6 occupied, 8 virtual), H$_2$O has 14 HF orbitals (10 occupied, 4 virtual). In the multi-qubit register, we assign each qubit to a spin-orbital, thus LiH (BeH$_2$ and H$_2$O) is (are) represented by 12 (14) qubits. We used the Quantum Learning Machine\cite{haidar2022open} (QLM) simulator that is able to simulate up to 40 qubits (with large memory), and we used the noiseless simulation mode (i.e. such ideal simulator mimics the behavior of performing the experiment with an infinite number of shots) in order to minimize computational run time and memory demands. 
 First, we tested the two methods on the LiH molecule at two bond lengths: $R = 1.0$ \r{A}  and $R = 3.0$ \r{A}. As shown in Table \ref{TableLiH}, we observe that UCCSD and sym-UCCSD brings similar numerical results. A similar observation can be made for the UCCSDT and sym-UCCSDT results. This demonstrates that the presence of spin and orbital symmetries in sym-UCC-VQE reaches the same result as UCC-VQE but in a more efficient way and without influencing the overall accuracy. 
 For example, Table \ref{TableSYMMETRY} shows that the number of parameters is reduced from 188 to 58 for the case of LiH, 
 which in practice accelerates the optimization 
 due to the fact that a lower dimensional optimization problem is tackled. We confirm the convergence of sym-UCCSD and sym-UCCSDT energies at gradient norm equal to 10$^{-10}$ inside BFGS optimizer  (see the convergence as a function of optimization steps in Figure 1 of the Supplementary Materials\cite{supplementary}).  \\\\

\begin{table}[h]
\begin{tabular}{|c|c|c|c|c|c|}
\hline
\multicolumn{2}{|c|}{\multirow{2}{*}{\textbf{Method}}} & \multicolumn{2}{|c|}{\textbf{R=1.0 \AA}} & \multicolumn{2}{|c|}{\textbf{R=3.0 \AA}} \\\cline{3-6}
\multicolumn{2}{|c|}{} & \textbf{Ener. (Ha)} & \textbf{Err. (Ha)} & \textbf{Ener. (Ha)} & \textbf{Err. (Ha)} \\\hline
\multicolumn{2}{|c|}{UCCSD} & $\textbf{-7.7844}5508682$ & $5.19  \times  10^{-6}$ & $\textbf{-7.798}7523587$ & $9.08  \times 10^{-5}$ \\\hline
\multirow{3}{*} &Sym.UCCSD  & $\textbf{-7.7844}5509065$ & $5.18  \times 10^{-6}$ & $\textbf{-7.798}75240726$ & $9.07  \times 10^{-5}$ 
 \\\hline
\multicolumn{2}{|c|}{UCCSDT(this work)} & $\textbf{-7.7844602}5863$ & $2.14 \times  10^{-8}$ & $\textbf{-7.798843}08319$ & $7.63\times 10^{-8}$\\\hline
\multirow{3}{*} & Sym.UCCSDT  & $\textbf{-7.7844602}5863$ & $2.13 \times  10^{-8}$ & $\textbf{-7.798843}08642$ & $7.30  \times 10^{-8}$
 \\\hline
\multicolumn{2}{|c|}{FCI (STO-3G)} & \multicolumn{2}{|c|}{$-7.78446028003$} & \multicolumn{2}{|c|}{$-7.79884315950$} \\\hline

%
\end{tabular}
\caption{UCC-VQE simulations for LiH at various bond lengths: $R = 1.0$ \r{A} and $R = 3.0$  \r{A}. The obtained UCC-energies are in Hartree (Ha). The error (Err) is the difference between the estimated UCC-energy and the FCI (the reference result is given in the last row).}
\label{TableLiH}
\end{table}

As mentioned in the literature\cite{anand2022quantum,barkoutsos2018quantum,romero2018strategies,grimsley2019trotterized, haidar2022open}, testing the Trotterization effect in the UCC ansatz is important since, as seen in equation \eqref{tettor},  using $t > 1$ might in principle, decreases the Trotterization error. 
Besides Trotterization study, a different method called disentangled  UCC ansatz\cite{doi:10.1063/1.5133059} has been introduced, and in \cite{D0CP01707H} the ordering of operators in this ansatz has been discussed. In our case we are interested in focusing on Trotterized ansatz truncated at Triple excitations. To test this in our symmetric operators, we use the sym-UCCSD and sym-UCCSDT methods with different Trotter steps $t$. It should be noted that when we increased it beyond 1, the same ansatz (and associated parameters) was chosen first and then we extended it.  We observe in Table 2 of the Supplementary Materials\cite{supplementary}  that increasing $t$, from $t=1$ up to $t=15$ steps does not impact the results. A similar conclusion is observed in the sym-UCCSDT case when we increased $t$ from $t=1$ up to $t=3$ steps. We have thus carefully checked that most of the operators that are involved in the symmetric version actually do not commute (see Table 3, Supplementary Materials\cite{supplementary} ) so that -- a priori -- there should be, at least, a slight difference in the optimized discrete energies with different values of $t$, but also there should be a difference on the
coefficients in front of the operators (results not shown : the coefficients for $t$ are those obtained for $t=1$ divided by $t$). The reason we found is that -- a posteriori, i.e. after computation -- we verified that, in both cases the operators associated to non-negligible coefficients do commute (i.e. the coefficients in front of non commuting operators are negligible).
We do not see reasons for this but it needs to be further analyzed.

\begin{figure}
\centering
\begin{subfigure}{.5\textwidth}
  \centering
  \includegraphics[width=\linewidth]{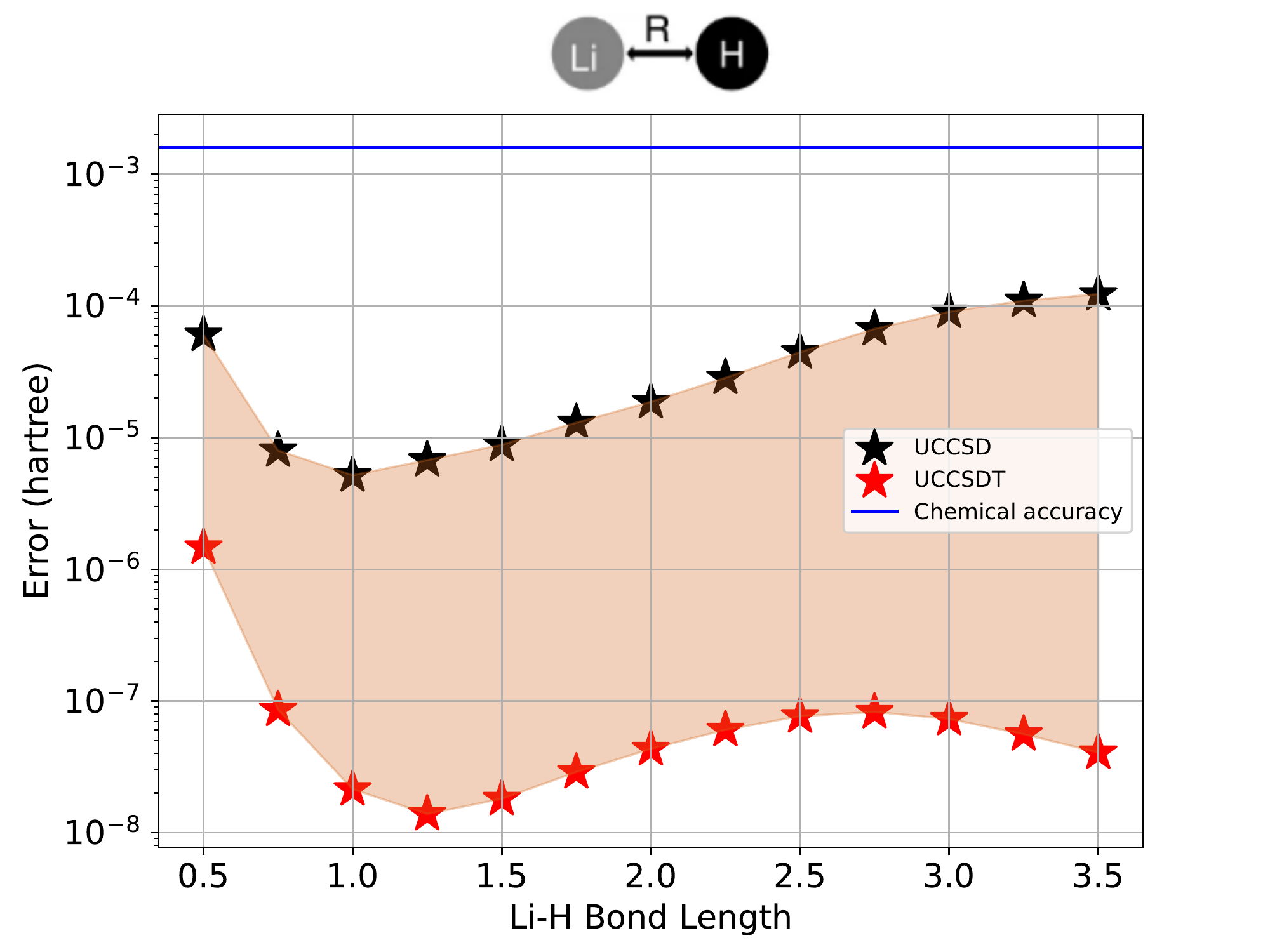}
  \caption{LiH: 12 qubits and  C$_{2v}$ point group }
  \label{fig:sub1}
\end{subfigure}%
\begin{subfigure}{.5\textwidth}
  \centering
  \includegraphics[width=\linewidth]{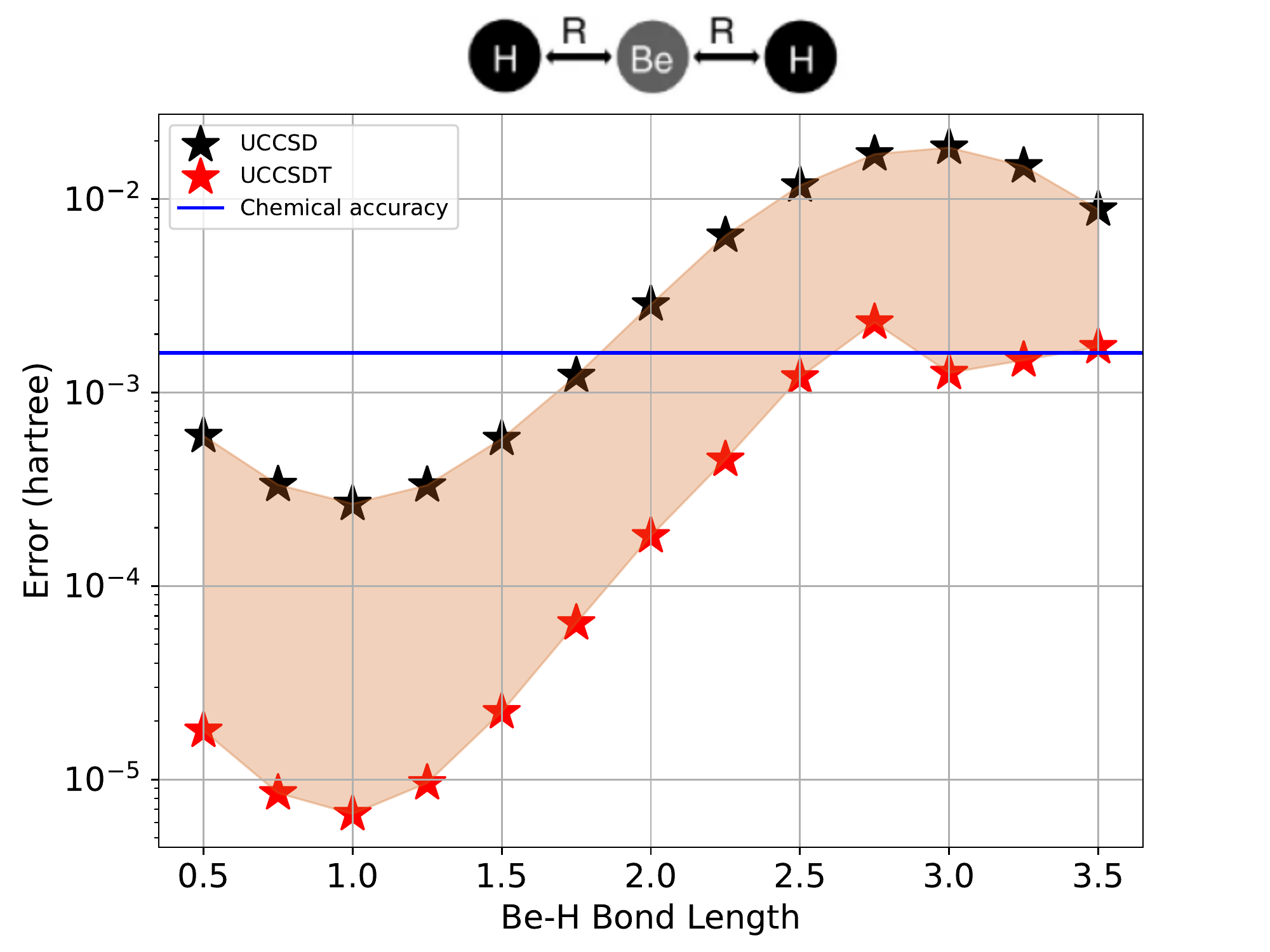}
  \caption{BeH$_2$: 14 qubits and D$_{2h}$ point group}
  \label{fig:sub2}
\end{subfigure}
\caption{Error of the UCCSD and UCCSDT simulations using both spin and point group symmetries for geometries LiH and BeH$_2$ at different bond lengths.  The horizontal blue line represents the chemical accuracy, i.e. 1.6 $\times 10^{-3}$ Ha.}
\label{fig:Lihandbeh2}
\end{figure}

In the remaining part of this Letter, we therefore use a UCC ansatz truncated at the Single, Double, and Triple excitations using a single Trotter step.\\
When applying the point group symmetry method, the number of optimization parameters was reduced between 14.2\% (case of BeH$_2$ having a D$_{2h}$ point group symmetry) and 30.8\% (case of LiH having a C$_{2v}$ point group symmetry), as shown in Table \ref{TableSYMMETRY}. This consequently will significantly reduce the gate cost of implementing triple excitations on quantum computers. Despite this reduction in parameters in the sym-UCCSDT circuit-ansatz, this circuit depth can be still considered as high compared to current NISQ devices capabilities. This is linked to the fact that we are using the CNOT staircase\cite{mcardle2020quantum,whitfield2011simulation} method that maps the exponential of each fermionic excitation into one- and two-qubit gates. Therefore, the number
of CNOT (2-qubit gate) increases with the level of excitations: 12  for single, 48 for double and 320 for Triples. Then the 
circuit depths are 21, 84, 464, respectively. This can be illustrated for example with the  BeH$_2$ case, as shown in Table \ref{TableSYMMETRY}. It contains 92 parameters, associated to the following excitations:  8 singles, 26 doubles, 52 triples, leading to a total CNOT gates in Sym-UCCSDT circuit equal to  12$\times$ 8 + 48 $\times$ 26 + 320 $\times$ 52 = 17984 
. Attempts for reducing the CNOT counts in triple excitation operators and higher-orders have been recently proposed \cite{magoulas2022cnot}. However, even with this reduction in CNOT counts, UCCSDT ansatz are not yet amendable to current NISQ devices.
\begin{table}[h!]
\begin{tabular}{cccccccc}
\hhline{========}
& Qbits & Sym. & Para Bef.(Af.) & $\%$ & $\Delta E_{SymUCCSD}$ \cite{cao2021towards} & $\Delta E_{SymUCCSD}^*$ & $\Delta E_{SymUCCSDT}^*$\\\hline\\
LiH& 12 & C$_{2v}$ & 188 (58) & 30.8\% & $1.09 \times 10^{-5}$ & $1.06 \times 10^{-5}$ & $2.16 \times 10^{-8}$ \\\\
H$_2$O & 14 & C$_{2v}$ & 340 (104) & 30.5\% & $1.09\times 10^{-4}$ & $1.0\times 10^{-4}$ & $2.11 \times 10^{-6}$\\\\
 BeH$_2$  & 14 & D$_{2h}$ & 644 (92) & 14.2\% & $3.82 \times 10^{-4}$ & $3.81 \times 10^{-4}$ & $6.63 \times 10^{-6}$\\
\hhline{========}
\end{tabular}
\caption{UCC-VQE simulations for LiH, H$_2$O and BeH$_2$ calculated at equilibrium. The geometry data are taken from CCCBDB-NIST Database\cite{DATAbase}. Columns 2 and 3 corresponds to the number of qubits and to the point group used in each of the molecules, respectively. The original parameters (and the parameters
after reduction) are given in the 4th column and the percent of the operators used (i.e single, double and triple excitations) after parameters reduction are in column 5. The UCC-VQE energy differences compared with the FCI energy (Ha) are shown from the 6$^{th}$ to 8$^{th}$ columns. Data of  $\Delta E_{SymUCCSD}$  
 are taken from ref\cite{cao2021towards}. Symbol(*) represents this work.}
\label{TableSYMMETRY}
\end{table}
Now let us discuss about sym-UCCSD and sym-UCCSDT energy errors, one of the main focus in the article. It is important to note that we will not consider the ordering of operators in the ansatz since it is believed to bring only little accuracy improvements.\cite{grimsley2019trotterized} However, it will, of course, remain of  potential interest to further study its impact. We believe that the inclusion of triple excitations should generate a sufficient improvement of the ansatz quality.
The energy error we study is the difference between the predicted UCC energy and FCI energy at the STO-3G basis set level.  When the molecules are chosen at their equilibrium geometry (see Table \ref{TableSYMMETRY}), the obtained sym-UCCSD error is similar to the one obtained in reference \cite{cao2021towards}, with only a very small difference of $\approx$ 10$^{-7}$ Ha. The calculated sym-UCCSDT improves the accuracy with respect to sym-UCCSD by about 3 orders of magnitude in LiH and by about 2-order of magnitudes in H$_2$O and BeH$_2$ molecules. This illustrates the capabilities of Triple excitations to recover the correlation energy missed by the sym-UCCSD ansatz at equilibrium. 
Moreover we calculate the sym-UCCSD and sym-UCCSDT energies for LiH and BeH$_2$ at several bond lengths $R$ ranging between 0.5   \r{A} to 3.5  \r{A}, see Figure \ref{fig:Lihandbeh2}. As shown in the Figure, in the LiH case, the sym-UCCSD and sym-UCCSDT errors remain  below the chemical accuracy for all bond lengths and the difference between sym-UCCSDT and sym-UCCSD error is  again about 3 orders of magnitude smaller. In the BeH$_2$ case, when R is 2.0  \r{A}, sym-UCCSD error starts to get above the chemical accuracy (1.6 $\times$ 10$^{-3}$ Hartree) and it gets further away as R increases. Overall, in term of error, Sym-UCCSDT is better than sym-UCCSD ones by about 1-2 order of magnitudes at all bond lengths. It remains below the chemical accuracy, except when R is 2.5  \r{A} and beyond, where the error starts to approach the blue line delimiting the chemical accuracy.  
When R is equal to 2.75 \r{A}, a discontinuity is observed. In fact  this has been also seen in a previous work, see reference  \cite{van2015polynomial}, around this bond length in BeH$_2$ calculated at STO-3G basis set level. There is one hypotheses which illustrate this anomaly. Indeed, 
 the importance of dynamic correlation starts to increase when the bond length is around 2.7\r{A}. At this point, according to figure 5 in \cite{van2015polynomial} the occupation
number of the natural orbitals starts to differ from zero or two. Hence many natural orbitals would have significant
occupation numbers (i.e different from zero), so the difference between
occupied and virtual orbitals tend to disappear. This probably leads to the inability of our sym-UCCSDT approach to capture some important excitations that might contribute strongly to the dynamical correlation, and thus the sym-UCCSDT errors could increase.
In general, this limitation from triple excitations at large bond lengths in BeH$_2$ can be probably improved by adding higher-order excitations but is consistent with classical couple cluster behaviours and the limits of single determinant approaches. The data  which were used to plot Figure \ref{fig:Lihandbeh2} are provided in Tables 4 and 5 from the Supplementary Informations\cite{supplementary}.
In order to compare the classical methods with UCC-VQE simulations, we calculated  CCSD, CCSD(T) and CCSDT-full errors for LiH, H$_2$O and BeH$_2$ at equilibrium. 
The results are shown  in Table \ref{TableClassicalcom}.   CCSD and sym-UCCSD errors are very similar. This is also observed in \cite{mizukami2020orbital} where Orbital optimized (OO)-UCCSD method has been tested. 
The
 CCSD(T) errors deviates clearly from sym-UCCSDT errors by at least one order of magnitudes. 
sym-UCCSDT also surpasses CCSDT-full(i.e Coupled cluster with a full treatment of singles,doubles and triples) in the three molecules, which is the most remarkable outcome here (see also same behavior in Figures 2 (a) and 2(b) of the supplementary materials\cite{supplementary} at some bond lengths in LiH and BeH$_2$, respectively). Such behaviour has been noticed in \cite{kohn2022capabilities}, where it is always CCSDT that has lower quality than UCCSDT, which was calculated analytically.

\begin{table}[h!]
\begin{tabular}{ccccccc}
\hhline{=======}
& $\Delta E_{CCSD}$ & $\Delta E_{SymUCCSD}^*$ & $\Delta E_{CCSD(T)}$ & $\Delta E_{CCSDT-full}$ & $\Delta E_{SymUCCSDT}^*$ & FCI\\\hline\\
LiH     & $1.05 \times 10^{-5}$ & $1.06 \times 10^{-5}$ & $2.11 \times 10^{-6} $ & $2.79 \times 10^{-8}$ & $2.16 \times 10^{-8}$ & $-7.882403410335502$\\\\
H$_2$O  & $1.17 \times 10^{-4}$ & $1.0  \times 10^{-4}$ & $4.91 \times 10^{-5}$ & $2.17 \times 10^{-5}$ & $2.11 \times 10^{-6}$ & $-75.01257824109094$\\\\
BeH$_2$ & $3.94 \times 10^{-4}$ & $3.81 \times 10^{-4}$ & $1.84 \times 10^{-4}$ & $4.22 \times 10^{-5}$ & $6.63 \times 10^{-6}$ & $-15.595176868923053$\\
\hhline{=======}
\end{tabular}
\caption{Error comparison between Classical methods (CCSD, CCSD(T) and CCSDT-full) and UCC-VQE (sym.UCCSD and sym.UCCSDT) simulations for LiH, H$_2$O and BeH$_2$ calculated at equilibrium. The geometry
data are taken from CCCBDB-NIST Database. The UCC-VQE results are referred from Table 2, and those of CCSD, CCSD(T), CCSDT-full are from Table 5 of the Supplementary Material\cite{supplementary}. The last column represents the FCI energies at STO-3G basis set. The error is the difference between the obtained energy and FCI in (Ha).}
\label{TableClassicalcom}
\end{table}




\section{Conclusion}
In conclusion, in this article, we presented the Trotterized UCCSDT ansatz. We simplified  the unitary coupled cluster triple excitation equations by using the spin and point group symmetries. By testing several molecules on QLM simulator, we demonstrated that the efficiency of the symmetric UCCSDT method can greatly improved. Indeed, reducing the non-important single, double and triple excitations from the circuit ansatz, one can accelerate the convergence process as well as reducing the circuit depth. Moreover, by performing several numerical simulations, we showed that the symmetric UCCSDT is able to reach a superior accuracy compared  to UCCSD (by at least two orders of magnitudes). 
In the future, as quantum simulators and Quantum Processing Units improve and enable to handle more gates, the Trotterized UCCSDT should enable to perform accurate calculations with large chemical basis sets (i.e beyond the minimal STO-3G). Overall, the Trotterized UCCSDT can bring  interesting results for the general quantum chemistry community since it is shown to be competitive with "gold-standard" CCSD(T) classical methods.

\begin{acknowledgement}
M.R. acknowledges funding from European Union's Horizon 2020 research and innovation program, more specifically the $\langle$NE$|$AS$|$QC$\rangle$ project under grant agreement No. 951821.
This work has also received funding from the European Research Council (ERC) under the European Union's Horizon 2020 research and innovation program (grant agreement No 810367), project EMC2 (JPP, YM).
Support from the PEPR EPiQ and HQI programs is acknowledged.

\end{acknowledgement}

\begin{suppinfo}
In the Supplementary Materials\cite{supplementary}, we present  all
data related to the computed total energies for sym-UCCSD and sym-UCCSDT found in Tables 1, 2 and 3. 
 Additionally, a detailed sketch algorithm that describe the triple excitations  within spin and point group symmetries is 
 presented and coded in our openVQE package\cite{openvqe}.
\end{suppinfo}

\bibliography{achemso-demo}

\end{document}